\documentclass[aps,twocolumn,prd,showpacs,showkeys,preprintnumbers,superscriptaddress,nobibnotes,floatfix,longbibliography,notitlepage,nofootinbib]{revtex4-1}

\pdfoutput=1
\usepackage{amsmath}
\usepackage{amsfonts}
\usepackage{amssymb}
\usepackage{mathrsfs}\usepackage{cancel}
\usepackage{accents}
\usepackage{mciteplus,slashed}
\usepackage{amssymb,cancel,amsmath,relsize}
\usepackage{mathrsfs} 
\usepackage{dcolumn} 
\usepackage{bm} 
\usepackage[caption=false]{subfig} 
\usepackage{appendix}
\usepackage{physics}
\usepackage{feynmp-auto}
\usepackage[T1]{fontenc}	
\usepackage{csvsimple}
\usepackage{hyperref}
\usepackage[capitalise]{cleveref}
\usepackage{booktabs}
\usepackage{graphicx}
\usepackage{mathrsfs}
\usepackage[utf8]{inputenc}
\usepackage[T1]{fontenc}
\graphicspath{{}}

\hypersetup{colorlinks,citecolor= blue,linkcolor= blue, urlcolor=blue}
 
\usepackage[dvipsnames]{xcolor}
\usepackage[normalem]{ulem}

\begin{document}

\preprint{IPMU22-0040}
\preprint{KEK-TH-2444}
\preprint{KEK-Cosmo-0293}
\preprint{KEK-QUP-2023-0031}

\title{Revealing Dark Matter Dress of Primordial Black Holes by Cosmological Lensing}

\author{Masamune Oguri}
\email{masamune.oguri@chiba-u.jp}
\affiliation{Center for Frontier Science, Chiba University, 1-33 Yayoi-cho, Inage-ku, Chiba 263-8522, Japan
}
\affiliation{Department of Physics, Graduate School of Science, \\
Chiba University, 1-33 Yayoi-Cho, Inage-Ku, Chiba 263-8522, Japan}

\author{Volodymyr Takhistov} 
\email{vtakhist@post.kek.jp}
\affiliation{International Center for Quantum-field Measurement Systems for Studies of the Universe and Particles (QUP, WPI),
High Energy Accelerator Research Organization (KEK), Oho 1-1, Tsukuba, Ibaraki 305-0801, Japan}
\affiliation{Theory Center, Institute of Particle and Nuclear Studies (IPNS), High Energy Accelerator Research Organization (KEK), Tsukuba 305-0801, Japan
}
\affiliation{The Graduate University for Advanced Studies (SOKENDAI), 1-1 Oho, Tsukuba, Ibaraki 305-0801, Japan}
\affiliation{Kavli Institute for the Physics and Mathematics of the Universe (WPI), UTIAS \\The University of Tokyo, Kashiwa, Chiba 277-8583, Japan}

\author{Kazunori Kohri}
\email{kohri-at-post.kek.jp}
\affiliation{Theory Center, Institute of Particle and Nuclear Studies (IPNS), High Energy Accelerator Research Organization (KEK), Tsukuba 305-0801, Japan
}
\affiliation{The Graduate University for Advanced Studies (SOKENDAI), 1-1 Oho, Tsukuba, Ibaraki 305-0801, Japan}
\affiliation{International Center for Quantum-field Measurement Systems for Studies of the Universe and Particles (QUP, WPI),
High Energy Accelerator Research Organization (KEK), Oho 1-1, Tsukuba, Ibaraki 305-0801, Japan}
\affiliation{Kavli Institute for the Physics and Mathematics of the Universe (WPI), UTIAS \\The University of Tokyo, Kashiwa, Chiba 277-8583, Japan}
\affiliation{Division of Science, National Astronomical Observatory of Japan (NAOJ), 2-21-1 Osawa, Mitaka, Tokyo 181-8588, Japan}

\date{\today}

\begin{abstract} 
Stellar-mass primordial black holes (PBHs) from the early Universe can directly contribute to the gravitational wave (GW) events observed by LIGO, but can only comprise a subdominant component of the dark matter (DM). The primary DM constituent will generically form massive halos around seeding stellar-mass PBHs. We demonstrate that gravitational lensing of sources at cosmological  ($\gtrsim$Gpc)  distances can directly explore DM halo dresses engulfing PBHs, challenging for lensing of local sources in the vicinity of Milky Way. Strong lensing analysis of fast radio bursts detected by CHIME survey already starts to probe parameter space of dressed stellar-mass PBHs,  and upcoming searches can efficiently explore dressed PBHs over $\sim 10-10^5 M_{\odot}$ mass-range and provide a stringent test of the PBH scenario for the GW events.  Our findings establish a general test for a broad class of DM models with stellar-mass PBHs, including those where QCD axions or WIMP-like particles comprise predominant DM. The results open a new route for exploring dressed PBHs with various types of lensing events at cosmological distances, such as supernovae and caustic crossings.
\end{abstract}
\maketitle

\section{Introduction}

The predominant form of matter in the Universe, the dark matter (DM), has been detected only through gravitational interactions and its nature remains mysterious~(e.g.~\cite{Gelmini:2015zpa}). The existence of black holes and their principal role in astronomy have been definitively established, including observations of supermassive black holes residing within galactic centers~(e.g.~\cite{Ghez:1998ph,Gillessen:2008qv,EventHorizonTelescope:2019dse,EventHorizonTelescope:2022xnr}). Primordial black holes (PBHs) formed the early Universe  prior to galaxies and stars could constitute all or a fraction of the DM abundance~(e.g.~\cite{Zeldovich:1967,Hawking:1971ei,Carr:1974nx,Meszaros:1975ef,Carr:1975qj,GarciaBellido:1996qt,Kawasaki:1997ju,Kohri:2007qn,Khlopov:2008qy,Frampton:2010sw,Bird:2016dcv,Kawasaki:2016pql,Inomata:2016rbd,Pi:2017gih,Inomata:2017okj,Garcia-Bellido:2017aan,Georg:2017mqk,Kocsis:2017yty,Ando:2017veq,Cotner:2016cvr,Cotner:2019ykd,Cotner:2018vug,Sasaki:2018dmp,Carr:2018rid,Kohri:2018qtx,Flores:2020drq,Deng:2017uwc,Kusenko:2020pcg,Carr:2020gox,Green:2020jor}).
Depending on the formation, PBHs can span decades in orders of magnitude in the mass range. 

PBHs in the stellar ($\sim 10-10^{2}M_{\odot}$) mass range are particularly intriguing as they have been intimately linked with the breakthrough observations of gravitational waves 
(GWs)~(e.g.~\cite{LIGOScientific:2016aoc}) by the LIGO-Virgo-KAGRA
collaboration (LVKC), which already detected dozens of stellar-mass BH binary merger events~\cite{LIGOScientific:2021usb,LIGOScientific:2021djp}. Such PBHs could contribute a sizable subdominant fraction of the DM energy density $f_{\rm PBH} = \Omega_{\rm PBH}/\Omega_{\rm DM}$ (e.g. \cite{Ali-Haimoud:2016mbv,Kohri:2018qtx,Serpico:2020ehh, Lu:2020bmd, Takhistov:2021aqx,Takhistov:2021upb}), with GW data indicating $f_{
\rm PBH} \sim \mathcal{O}(10^{-3})$ (e.g.~\cite{Bird:2016dcv,Clesse:2016vqa,Sasaki:2016jop,Franciolini:2021tla}).

Constituting overdensities in an expanding Universe, PBHs will seed the growth of DM halos~\cite{Mack:2006gz,Ricotti:2007au} following the theory of spherical gravitational collapse~\cite{Bertschinger:1985pd}. Hence, stellar-mass PBHs forming a subdominant DM component will be generically engulfed in massive halo dresses composed of the primary DM component.

A well motivated possibility is that DM is predominantly composed of  axions, which can naturally arise from fundamental theory~\cite{Svrcek:2006yi} and directly connected with the strong CP problem~\cite{Peccei:1977ur,Weinberg:1977ma,Wilczek:1977pj,Dine:1981rt}. While significant experimental efforts are underway (see \cite{Adams:2022pbo} for review), axions remain elusive. 
In general, DM halos can be composed of a slew of motivated DM candidates~(see \cite{Bertone:2004pz} for review), including also weakly interactive massive particles (WIMP-like particles) with typical mass in the GeV to 100~TeV range that often arise in models addressing the hierarchy problem.
With DM concentrations significantly exceeding that of ambient cold DM, dark halo dresses surrounding PBHs make for an excellent laboratory to explore DM beyond conventional techniques.

In this work we propose cosmological lensing of fast radio bursts (FRBs) as a novel direct probe for exploring dark halo dresses around PBHs and a way to distinguish them with bare PBHs, allowing for unique tests of a wide range of DM models where stellar-mass PBHs can contribute to the detected GW events. 
In contrast to some of the previously discussed signatures associated with specific DM models comprising dark halos around PBHs (e.g.~\cite{Hertzberg:2020kpm,Nurmi:2021xds}), our analysis establishes a general direct test of dressed PBHs in models where QCD axions and WIMP-like particles could be the primary DM contributors, among others.
 
\begin{figure*}[t]
    \centering
    \includegraphics[trim={0cm 0cm 0 0},clip,width=0.65\textwidth]{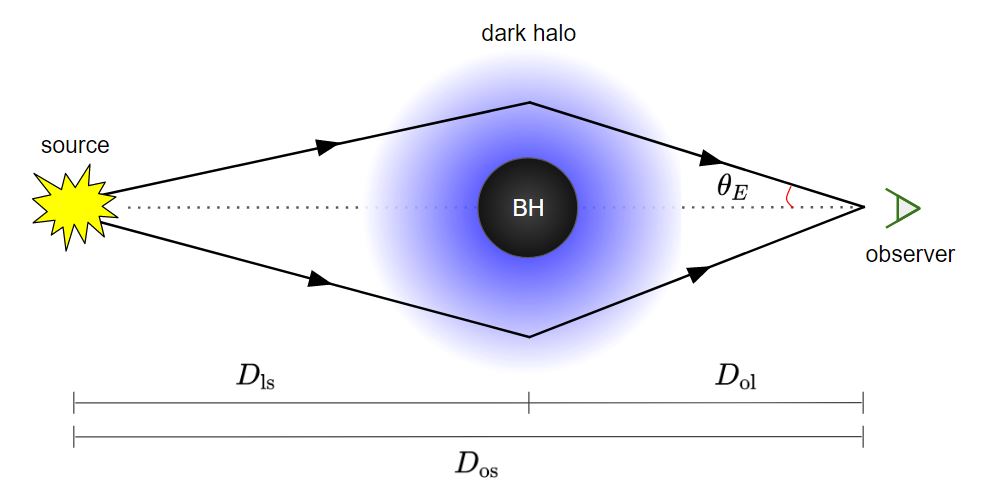}
    \caption{Schematic depicting cosmological lensing of a source by a PBH dressed in a DM halo.}
    \label{fig:schematic}
\end{figure*}

\section{Dark Matter Halo Formation}

Stellar-mass PBHs will generically seed DM halos from accretion of surrounding smooth background of the primary DM component, until dressed PBHs reside within galactic halos.  
The halo surrounding a PBH will grow of order unity in mass during radiation-dominated era, and as $\propto (1+z)^{-1}$ during matter-dominated era at redshift $z$. Thus, following numerical and analytic calculations the dark PBH halo grows with redshift as
\cite{Mack:2006gz,Ricotti:2007au}
\begin{align} \label{eq:haloparam}
M_{\rm h}(z_{\rm c}) =&~ 3 \Big(\dfrac{1000}{1+z_{\rm c}}\Big) M_{\rm PBH} \\
R_{\rm h}(z_{\rm c}) =&~ 0.019~{\rm pc} \Big(\dfrac{M_{\rm h}}{M_{\odot}}\Big)^{1/3} \Big(\dfrac{1000}{1+z_{\rm c}}\Big)~.   \notag
\end{align}
The description of halo growth holds until $z_{\rm c} \sim 30$, when PBHs start to interact with nonlinear cosmic structures. We note that the high density of DM in halos surrounding PBHs protects them against tidal stripping. These effects are not expected to be significant unless PBHs are orbiting very close to center of galaxy~\cite{Nurmi:2021xds}.
 
The resulting dark halo develops a density profile  $\rho \propto r^{-9/4}$~\cite{Bertschinger:1985pd,Berezinsky:2013fxa}, as confirmed by $N$-body simulations~\cite{Adamek:2019gns,Serpico:2020ehh,Boudaud:2021irr}. From Eq.~\eqref{eq:haloparam} and parametrization
\begin{equation}
    \rho_{\rm h} (r) = \rho_0 \Big(\dfrac{R_{\rm h}}{r}\Big)^{9/4}
\end{equation}
we find $\rho_0 \simeq 8701 (1000/(1+z_{\rm c}))^{-3}~M_\odot~{\rm pc}^{-3}$. 
Here we consider the characteristic general case that DM is collisionless. Depending on assumptions and DM model, the dark halo profile could have behavior distinct from $\sim r^{-9/4}$ at inner radii~(e.g.~\cite{Carr:2020mqm,Feng:2021qkj,Boudaud:2021irr}). However, these effects appear at the innermost radii where the enclosed dark halo mass is comparable to that of seeding PBH~\cite{Adamek:2019gns}. Hence, we do not expect that this will significantly affect our results. Our analysis can be readily extended to accommodate these effects and we leave their detailed evaluation in the context of specific scenarios for future work. Further, we do not focus here on WIMPs with significant annihilation channel or DM decaying on time-scales shorter than halo formation that due to increased density in DM halo have already been constrained by indirect-detection observations such as gamma-rays~(e.g.~\cite{Lacki:2010zf}).

\section{PBH-halo Lensing}

We now estimate the impact of the surrounding dark halo on PBH lensing. A central quantity characterizing gravitational lensing is the surface mass density $\Sigma$, which is the density profile of the lens projected along the line-of-sight. The corresponding dimensionless quantity is called convergence $\kappa=\Sigma/\Sigma_{\rm cr}$, with critical surface density being~(e.g.~\cite{Narayan:1996ba})
$\Sigma_{\rm cr} = (1/4\pi G) (D_{\rm os}/(D_{\rm ol} D_{\rm ls}))~$, where we take  $G$ to be the gravitational constant, $D_{\rm os}$ is the angular diameter distance between the observer and the source, $D_{\rm ol}$ between the observer and the lens, $D_{\rm ls}$ between the lens and the source. In Fig.~\ref{fig:schematic} we schematically illustrate gravitational lensing by a dressed PBH. Throughout, we assume natural units, $c = 1$.
The critical surface density corresponds to the
mean surface density enclosed within the Einstein radius $r_{\rm E}$ related to Einstein angle as $r_E = \theta_E D_{\rm ol}$ (tangential critical radius, defining the size of the Einstein ring), with enclosed mass being $M(<r_{\rm E}) = \pi \Sigma_{\rm cr} r_{\rm E}^2$.

For the dark halo around a PBH we find its surface density to be
\begin{align} \label{eq:sigma}
    \Sigma_{\rm h} (r) =&~ \int_{-\infty}^{\infty} dz \rho_0 \Big(\dfrac{R_{\rm h}}{\sqrt{r^2+z^2}}\Big)^{9/4}  \notag
    \\
    =&~\rho_0 R_{\rm h} \sqrt{\pi} \dfrac{\Gamma(5/8)}{\Gamma(9/8)} \Big(\dfrac{R_{\rm h}}{r}\Big)^{5/4}~.
\end{align}
Here we have ignored the truncation of the density profile beyond $R_{\rm h}$, as well as the truncation at the inner boundary. While this leads to a slight overestimation of the effects of dark halo, we expect these effects to be insignificant as long as $r_E \ll R_{\rm h}$, which is indeed the case for our parameters of interest in this work (see Appendix). 
From Eq.~\eqref{eq:sigma}, we obtain the average halo surface density as
\begin{align} \label{eq:avsigma}
\bar{\Sigma}_{\rm h}(< r) =&~ \dfrac{1}{\pi r^2} \int_0^r dr \Sigma_{\rm h}(r) 2 \pi r \notag\\
=&~\dfrac{8 \rho_0 R_{\rm h}}{3}\sqrt{\pi}\dfrac{\Gamma(5/8)}{\Gamma(9/8)}\Big(\dfrac{R_{\rm h}}{r}\Big)^{5/4}~.
\end{align}
In contrast, the surface density of PBH with mass $M_{\rm PBH}$ can be modelled simply by $\bar{\Sigma}_{\rm PBH}(<r)=M_{\rm PBH}/(\pi r^2)$ assuming that its density profile is described by the Dirac's delta function at the origin.

We can now directly estimate the average convergence $\bar{\kappa}(<r)$, which for the Einstein radius satisfies $\bar{\kappa}(<r_E) = 1$ and delineates the regimes of strong and weak gravitational lensing~(e.g.~\cite{Narayan:1996ba}). From Eq.~\eqref{eq:avsigma} we obtain the average convergence for the halo $\bar{\kappa}_{\rm h}$ as well as the average convergence for PBH $\bar{\kappa}_{\rm PBH}$ to be
\begin{equation} \label{eq:halocon}
    \bar{\kappa}_{\rm h}(< r) = \dfrac{\bar{\Sigma}_{\rm h}(<r)}{\Sigma_{\rm cr}}~~~,~~~\bar{\kappa}_{\rm PBH}(< r) = \dfrac{M_{\rm PBH}}{\pi r^2 \Sigma_{\rm cr}}~.
\end{equation}
Comparing the average convergence of halo within PBH Einstein radius, i.e. $\bar{\kappa}_{\rm h}( < r_{E, {\rm PBH}}) = 1$, allows to estimate the effect of dark halo on PBH lensing.

Adding the contribution of the dark halo to a PBH, the new Einstein radius $r_{E, {\rm tot}}$ is found by computing
\begin{equation} \label{eq:eintot}
    \bar{\kappa}_{\rm PBH}(<  r_{E, {\rm tot}} ) + \bar{\kappa}_{\rm h} (<  r_{E, {\rm tot}} ) = 1~.
\end{equation}
We can now compare results of Eq.~\eqref{eq:eintot} to that of isolated PBH $r_{E, {\rm PBH}}$, which is typically employed for PBH microlensing searches~(e.g.~EROS~\cite{EROS-2:2006ryy}, OGLE~\cite{Niikura:2019kqi} and Subaru HSC~\cite{Niikura:2017zjd}). 

From the Einstein radius for a PBH with the dark halo, we can define the ``effective PBH mass'' $M_{\rm PBH}^{\rm eff}$ as 
\begin{equation} \label{eq:meff}
r_{E, {\rm tot}} (M_{\rm PBH}) =  r_{E, {\rm PBH}} (M_{\rm PBH}^{\rm eff})~ . 
\end{equation}
The Einstein radius of an isolated point mass lens with $M_{\rm PBH}^{\rm eff}$ matches the Einstein radius of a PBH with mass $M_{\rm PBH}$ with the dark halo. 
For a given PBH with mass $M_{\rm PBH}$, its the lensing cross section is on the order to $\pi \{r_{E, {\rm tot}}(M_{\rm PBH})\}^2$, and by using Eq.~\eqref{eq:meff} we can rewrite it as $\pi \{r_{E, {\rm PBH}}(M_{\rm PBH}^{\rm eff})\}^2=M_{\rm PBH}^{\rm eff}/\Sigma_{\rm cr}$, suggesting that a halo of a PBH with mass $M_{\rm PBH}$ enhances its lensing cross section by $M_{\rm PBH}^{\rm eff}/M_{\rm PBH}$.

\section{Cosmological Lensing with Fast Radio Bursts}

In order to efficiently distinguish between dressed and bare PBHs, for a given PBH mass the ratio of mass in Eq.~\eqref{eq:meff} should satisfy $M_{\rm PBH}^{\rm eff}/M_{\rm PBH} \gg 1$. As we explicitly demonstrate in Appendix, gravitational lensing of sources at cosmological distances with redshifts $z \sim \mathcal{O}(1)$ can efficiently reveal dark dresses of PBHs. On the other hand, as we verify, it is challenging to distinguish dressed and bare PBHs through gravitational lensing of local sources located at $z \ll 1$.

Particularly promising sources for cosmological lensing are FRBs (see \cite{Oguri:2019fix} for review).
These are luminous transient radio signals of millisecond time-scales at cosmological distances, able to provide favorable sources for lensing by bare stellar-mass PBHs~\cite{Munoz:2016tmg,Laha:2018zav,Liao:2020wae}.
Strong lensing of FRBs by stellar-mass PBHs will generate two distinct images of the bursts whose time delays are sufficiently large to resolve them in the time domain~\cite{Munoz:2016tmg}. 
More than several hundred FRBs, including some repeating, have been identified by the Canadian Hydrogen Intensity Mapping Experiments (CHIME) experiment~\cite{CHIMEFRB:2021srp}. Analysis of lensing of 536 FRBs~\cite{Krochek:2021opq} detected by CHIME, as well as 593 FRBs with cross-correlation~\cite{Zhou:2021ndx}, constrained stellar-mass PBH DM abundance at the level of $f_{\rm PBH} \sim \mathcal{O}(10^{-2})$.
Additional details for the framework of FRB strong lensing we employ can be found in Appendix. We note that the CHIME collaboration has already carried out an extremely detailed related lensing search and constraints involving interferometric
lensing, which yields time resolution of up to $\sim100$ nanoseconds~\cite{CHIMEFRB:2022xzl,Leung:2022vcx}.

With an extended halo, for dressed PBHs we can compute the lensing optical depth as
\begin{align}
\label{eq:halotau}
\tau^{\rm w/\,h}(M_{\rm PBH}) =&~ \int_0^{z_S}d\chi (z_L)(1+z_L)^2 \\
&~ \times n_{\rm PBH}(M_{\rm PBH}) \sigma (M_{\rm PBH}^{\rm eff}, z_L) \notag\\
=&~ \Big(\dfrac{M_{\rm PBH}^{\rm eff}}{M_{\rm PBH}}\Big) \Big(\dfrac{f_{\rm PBH}(M_{\rm PBH}^{\rm eff})}{f_{\rm PBH}(M_{\rm PBH})} \Big)^{-1}\tau^{\rm w/o\,h}(M_{\rm PBH}^{\rm eff})~, \notag
\end{align}
where $\chi(z)$ is the comoving distance at redshift $z$, $n_{\rm PBH} = f_{\rm PBH} \Omega_{\rm cdm} \rho_{\rm crit}/M_{\rm PBH}$ is the comoving number density of PBHs, with $f_{\rm PBH}$ being the fraction PBH DM abundance, $\rho_{\rm crit}$ is the critical density, $\Omega_{\rm cdm} = 0.24$ is the Universe's cold DM density and $\sigma$ is the lensing cross-section characterizing the lens and dependence on the impact parameters (see Appendix). The lensing optical depth $\tau^{\rm w/o\,h}$ denote the traditional PBH lensing cross section without the halo contribution as described in Appendix.
Optical depth for point source lensing by bare PBHs can be recovered from Eq.~\eqref{eq:halotau} by substituting $M_{\rm PBH}$ for {$M^{\rm eff}_{\rm PBH}$. Prefactors in the last expression in Eq.~\eqref{eq:halotau} originate from the fact that for bare PBH lensing, $n_{\rm PBH}$ remains the same while the lensing cross-section is modified. While here we derive the lensing cross-section assuming the point mass lens with mass $M^{\rm eff}_{\rm PBH}$,
the computation involving lensing cross section could be further refined by employing the point mass added to the model of dark halo mass distribution is possible, analysis of which we leave for future work~\cite{dressedPBH:fut}.
 
\begin{figure}[t]
    \centering
    \includegraphics[trim={0cm 0cm 0 0},clip,width=\columnwidth]{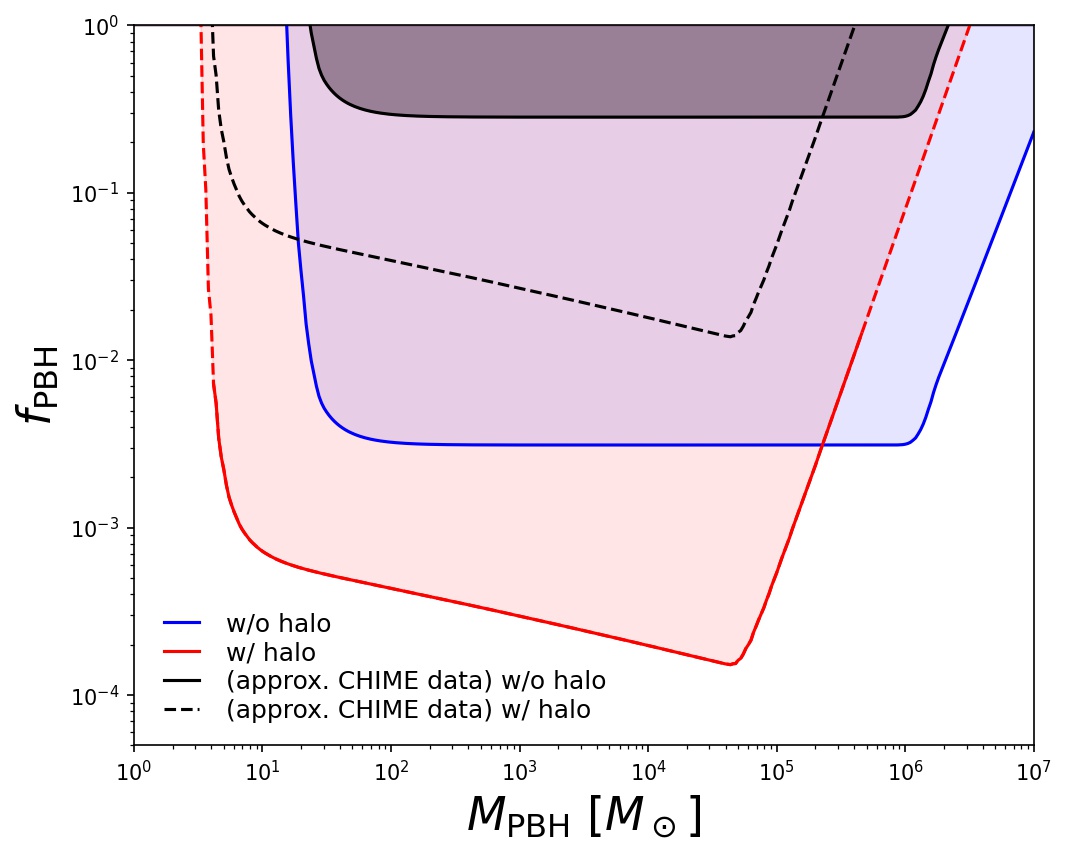}
    \caption{PBH DM parameter space as a function of bare PBH mass, depicting dressed and bare PBHs.
    Projections at 95\% confidence level for dressed (``w/ halo'') as well as bare (``w/o halo'') PBHs considering $5 \times 10^4$ upcoming FRB observations are shown. Regions of "approx. CHIME data" denote estimated limits from 550 FRBs, an approximate amount already collected by CHIME. Dashed lines depict regions above $f_{\rm PBH} = 10^{-2}$ where large halo dress accumulation around PBHs could be affected.
    }
    \label{fig:halopbhmain}
\end{figure}

For a given redshift-distribution function of $N(z)$ FRB sources, the integrated optical depth is determined by
\begin{equation}
\label{eq:tauint}
    \overline{\tau}(M_L) = \int dz \tau(z, M_L)N(z)~.
\end{equation}
For the optically thin regime $\overline{\tau} \ll 1$, relevant here, the lensing probability is $P_l = 1 - e^{-\overline{\tau}} \simeq \overline{\tau}$. Thus, for $N_o$ observed FRB events, the number of lensed events follows $N_o \overline{\tau}$. For $N(z)$, we adopt the constant comoving-density model with cutoff $z_{\rm cut} = 0.5$ as in Ref.~\cite{Munoz:2016tmg} (see Appendix).

Given that $\tau \propto f_{\rm PBH}$, non-detection of lensed FRBs by dressed PBHs from $N_o$ observations translates into $2\sigma$ (95\% confidence level) Poisson statistics of
\begin{equation} \label{eq:fpbhconst}
    f_{\rm PBH} (M_{\rm PBH}) < \Big(\dfrac{M_{\rm PBH}^{\rm eff}}{M_{\rm PBH}}\Big)^{-1} \dfrac{3.05}{N_o \tau^{\rm w/o\,h} (M_{\rm PBH}^{\rm eff}, f_{\rm PBH} =1 )}~.
\end{equation}
The constraint on dressed PBHs of Eq.~\eqref{eq:fpbhconst} translates into constraint on bare PBHs when $M_{\rm PBH}^{\rm eff} = M_{\rm PBH}$.
In Fig.~\ref{fig:halopbhmain} we depict our computed projections for non-detection of bare and dressed PBHs for upcoming detection of $5 \times 10^4$ FRB events. Unlike the case of bare PBHs, we stress that sensitivity to dressed PBHs significantly increases with mass due to growth of dark halo dress around PBHs.
We adopt the threshold of the flux ratio for the lensing pair search to $5$, as in analysis of Ref.~\cite{Munoz:2016tmg}. We note that as the PBH mass increases, eventually the sensitivity becomes limited to due to maximum time delay in CHIME data (taken to be 1 minute)~\cite{CHIMEFRB:2021srp}. At lower mass range, below $\sim 10 M_{\odot}$, sensitivity decreases due to requirement that time delay is larger than critical value $\Delta \bar{t} = 1$~ms.
We have verified that assuming $\sim550$ FRBs our estimated limits
are in general agreement with dedicated detailed analyses of bare PBHs using CHIME catalogue of Ref.~\cite{Krochek:2021opq,Zhou:2021ndx}. Since above $f_{\rm PBH} \sim \mathcal{O}(10^{-2})$ dressed PBHs might not fully develop a DM halo, we depict these limits with dashed line and leave a more detailed study of this complication for future work~\cite{dressedPBH:fut}. As we demonstrate, with upcoming observations of $\sim 10^4$ FRB events, CHIME or other FRB surveys will be able to probe dressed PBHs in the region relevant for observed GWs and have the capability to definitely distinguish with bare PBHs. Put another way, if the PBH scenario for the GW events is correct, we should be able to detect FRB lensing events from observations of $\gtrsim 10^4$ FRBs, indicating that we may confirm or exclude the PBH scenario of GW events by the near-future FRB observations. Dressed PBHs are especially relevant in this region of interest, while the region expanding to the right in PBH mass-range is already under significant pressure from multiple constraints~(e.g.~\cite{Kohri:2014lza,Ali-Haimoud:2016mbv,Serpico:2020ehh, Lu:2020bmd, Takhistov:2021aqx,Takhistov:2021upb}).

Our findings establish novel opportunities for exploring directly models of dressed PBHs with dark halos through cosmological lensing, and can be applied to a variety of PBH lensing events at cosmological distances,  including caustic crossings~\cite{Oguri:2017ock} and Type Ia supernovae~\cite{Zumalacarregui:2017qqd}. The analysis of lensing of these sources involves additional complexity and a detailed study is a topic of separate work~\cite{dressedPBH:fut}.

\vspace{-0.5cm}
\section{Conclusions}

Stellar-mass PBHs contributing to the DM abundance have been intimately linked with the detected GW events by LIGO. However, such PBHs can comprise only a sub-dominant fraction of the DM and are expected to be dressed in DM halos composed of a distinct primary DM component, such as QCD axion or WIMP-like particles. We advance cosmological lensing as a novel general method for directly identifying dressed PBHs and distinguishing them with bare PBHs, challenging for lensing of local sources. As we have demonstrated, strong lensing of observed FRBs from CHIME already begin testing dressed PBHs when stellar-mass PBHs constitute at the sub-percent level to the DM abundance. Upcoming observations by CHIME, or other FRB surveys, can explore the parameter space associated with detected GW events. Our findings open a new route for studying dressed PBHs with other cosmological lensing sources, such as caustic crossings and supernovae.

\section*{Acknowledgements}

This work was supported in part by JSPS KAKENHI Grant Numbers
JP17H01131 (K.K.) and by MEXT KAKENHI Grant Numbers JP20H04750,
JP22H05270 (K.K.), JP20H05856 (M.O.), JP20H00181 (M.O.). 23K13109 (V.T.), V.T. was also supported by World Premier International Research Center Initiative (WPI), MEXT, Japan.

\appendix

\section{Lensing by Dressed PBHs for Cosmological and Local Sources}
\label{sec:lensing}

To obtain a favorable configuration for lensing where dressed PBHs can be efficiently detected and distinguished from bare PBHs, we analyze key lensing quantities for local as well cosmological sources. For local sources, we assume $D_{\rm ol}=100$~kpc and $D_{\rm os}=770$~kpc for microlensing of Andromeda M31 galaxy, as employed for analysis of PBHs by Subaru HSC~\cite{Niikura:2017zjd}. For cosmological sources, we consider location at $z_S = 1$ as characteristic of FRBs and PBHs located at $z_L = 0.5$.

In Fig.~\ref{fig:lenscomp_1}, we display the average convergence $\bar{\kappa}(<1)$ for bare PBHs as well as their surrounding dark halo computed following Eq.~\eqref{eq:halocon} of the main text. With $\bar{\kappa} = 1$ corresponding to the Einstein radius and delineating the boundary between weak and strong lensing, we find that for $M_{\rm PBH} \gtrsim \mathcal{O}(1) M_{\odot}$ strong lensing by cosmological sources can efficiently distinguish bare PBHs with halo-dressed PBHs. On the other hand, the contribution of the dark halo to $\bar{\kappa}(<r)$ is subdominant for local sources even at around the Einstein radius. 

We also display in Fig.~\ref{fig:lenscomp_2} computed Einstein radius with and without the dark halo to explicitly verify the effect of the dark halo.  We observe that, for cosmological lensing, dressed PBHs will have a significantly distinct Einstein radius compared to bare PBHs for $M_{\rm PBH} \gtrsim 10^{-2} M_{\odot}$. The difference of the Einstein radii are larger for higher PBH mass, indicating that PBH lensing with higher PBH masses can probe DM dress more efficiently. The figure also confirms that the halo size $R_{\rm h}$ is typically much larger than the Einstein radius. 

\begin{figure*}[h]
    \centering
    \includegraphics[trim={0cm 0cm 0 0},clip,width=1\columnwidth]{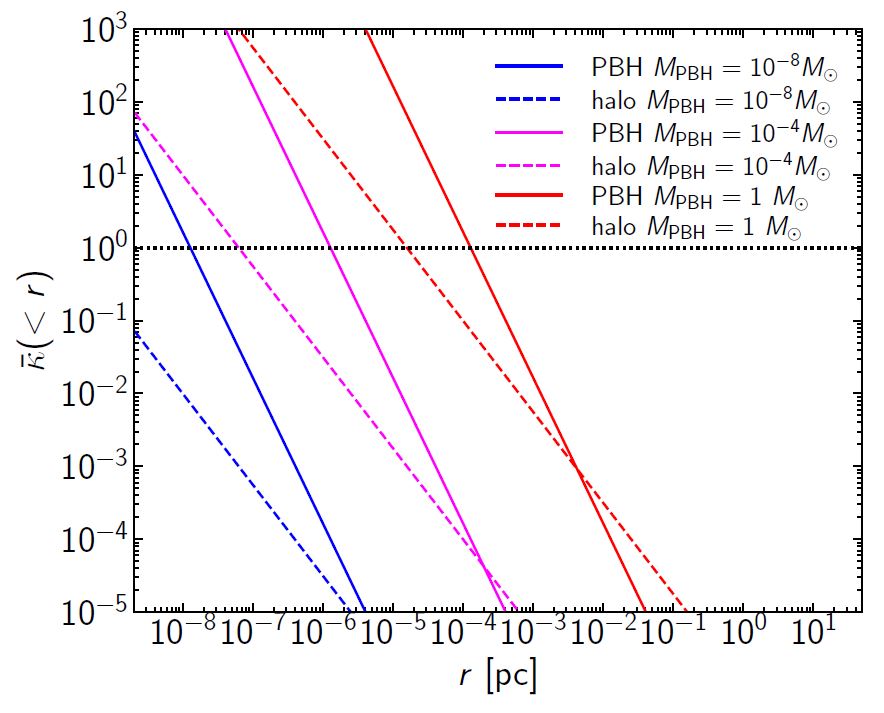}
    \hspace{1mm}
    \includegraphics[trim={0cm 0cm 0 0},clip,width=1\columnwidth]{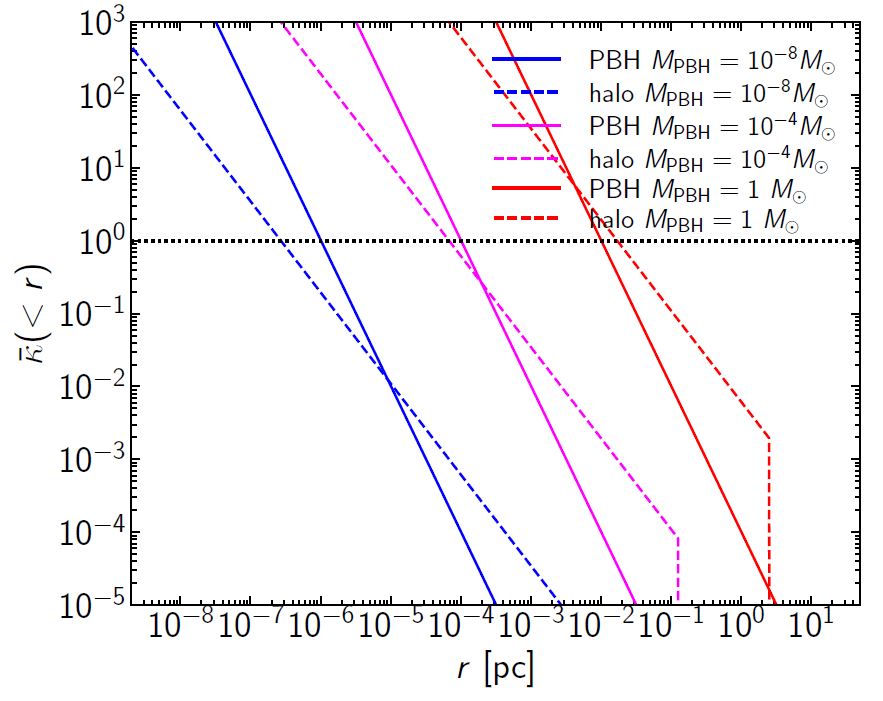}
    \caption{\textbf{Left:} Comparison of the average convergence for a PBH and its surrounding halo for three different PBH masses, in case of lensing of local sources at M31 Andromeda galaxy. The average convergence profiles of halos are truncated at $R_{\rm h}$ to indicate the halo size. The horizontal dotted line indicates $\bar{\kappa}(<r)=1$, corresponding to the Einstein radius. \textbf{Right:} Similar to the left panel, but for cosmological sources assuming $z_S = 1$ and $z_L = 0.5$.}
    \label{fig:lenscomp_1}
\end{figure*}

\begin{figure*}[h]
    \centering
    \includegraphics[trim={0cm 0cm 0 0},clip,width=1\columnwidth]{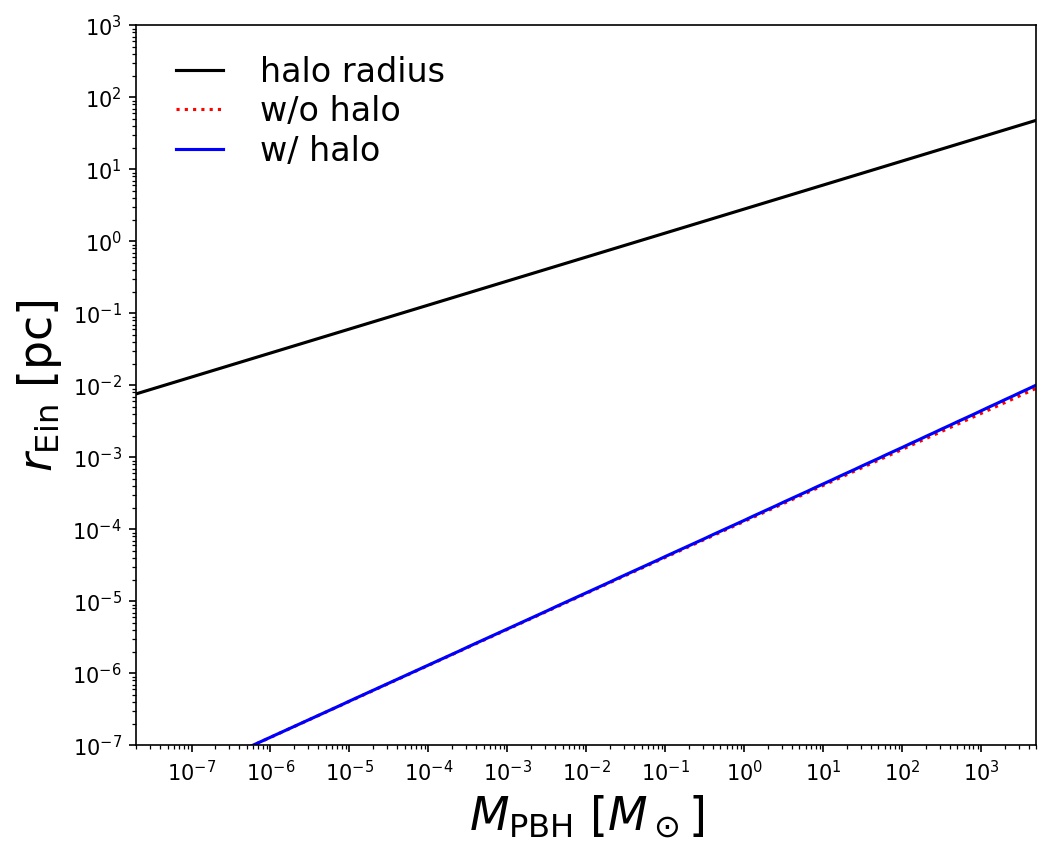}
    \includegraphics[trim={0cm 0cm 0 0},clip,width=1\columnwidth]{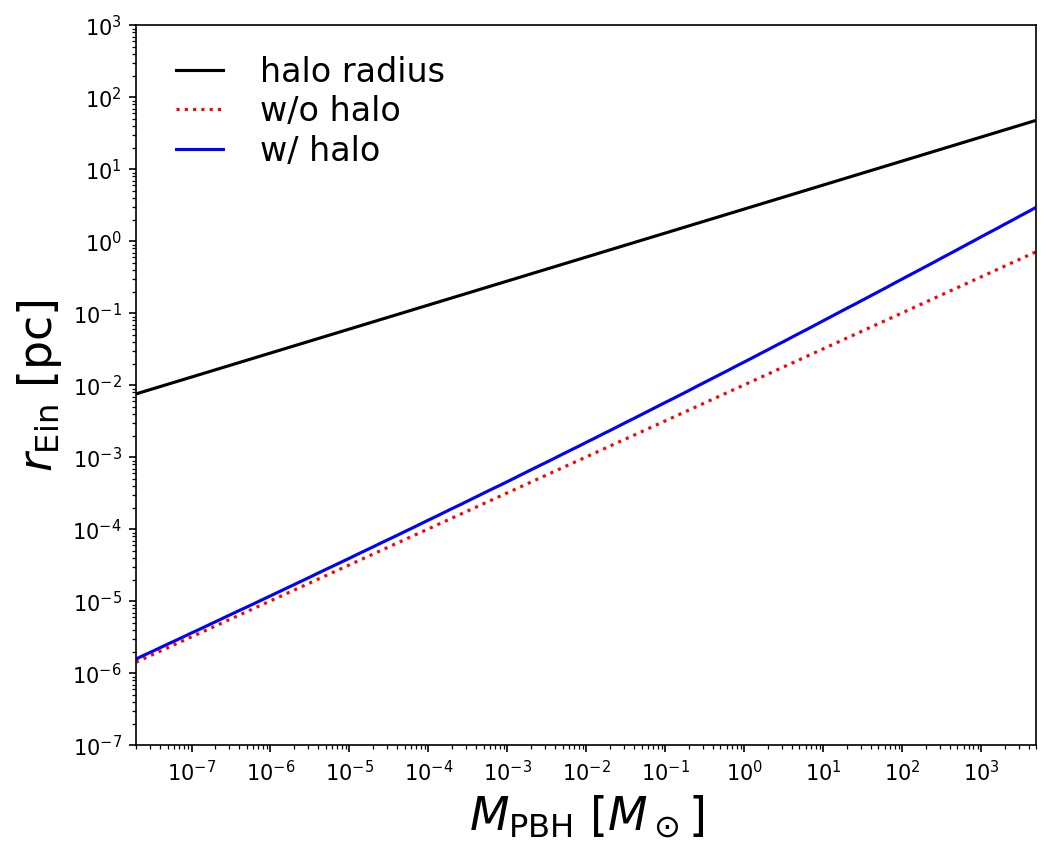}
    \caption{\textbf{Left:} Comparison of the Einstein radii with and without the dark halo as a function of the bare PBH mass $M_{\rm PBH}$, in case of lensing of local sources at M31 Andromeda galaxy.  \textbf{Right:} Similar to the left panel, but for cosmological sources assuming $z_S = 1$ and $z_L = 0.5$.}
    \label{fig:lenscomp_2}
\end{figure*}

Using Eq.~\eqref{eq:meff}, in Fig.~\ref{fig:halomeff} we compute effective PBH mass  $M_{\rm PBH}^{\rm eff}$ that reproduces the Einstein radius of dressed PBHs with mass $M_{\rm PBH}$, for lensing by local (M31) and cosmological sources. Again, we observe that, for lensing of cosmological sources, dressed PBHs can be readily distinguished from bare PBHs with $M_{\rm PBH}^{\rm eff}/M_{\rm PBH} \gg 1$ when $M_{\rm PBH} \gtrsim 10^{-2} M_{\odot}$. On other other hand, we confirm that probing DM dress is difficult for PBH lensing of local sources, as the difference between $M_{\rm PBH}^{\rm eff}$ and $M_{\rm PBH}$ is small even for PBHs with high $M_{\rm PBH}$.

\begin{figure*}[t]
    \centering
    \includegraphics[trim={0cm 0cm 0 0},clip,width=1\columnwidth]{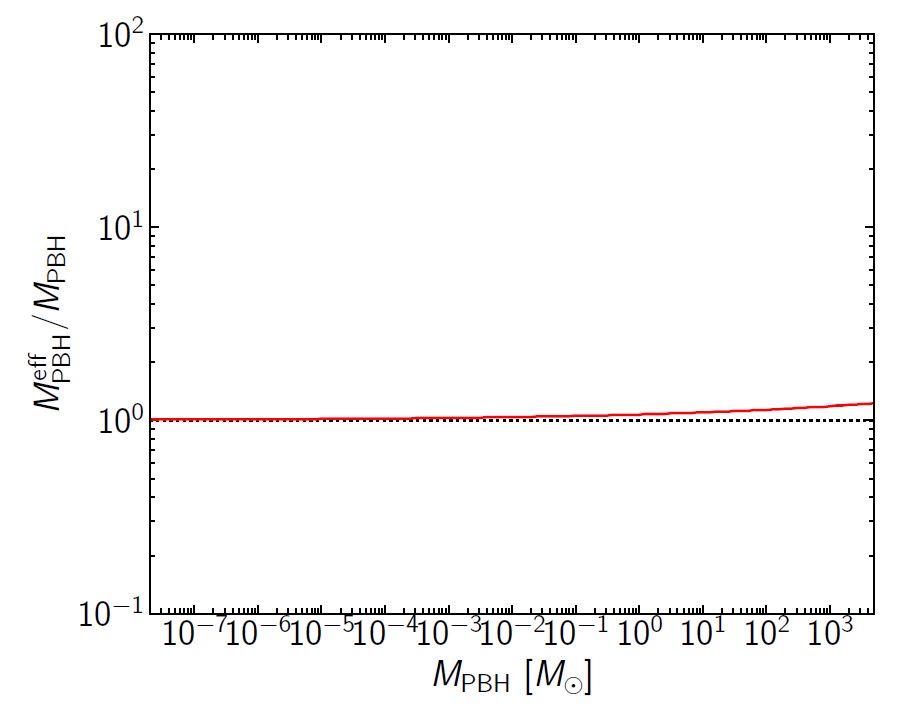}
    \includegraphics[trim={0cm 0cm 0 0},clip,width=1\columnwidth]{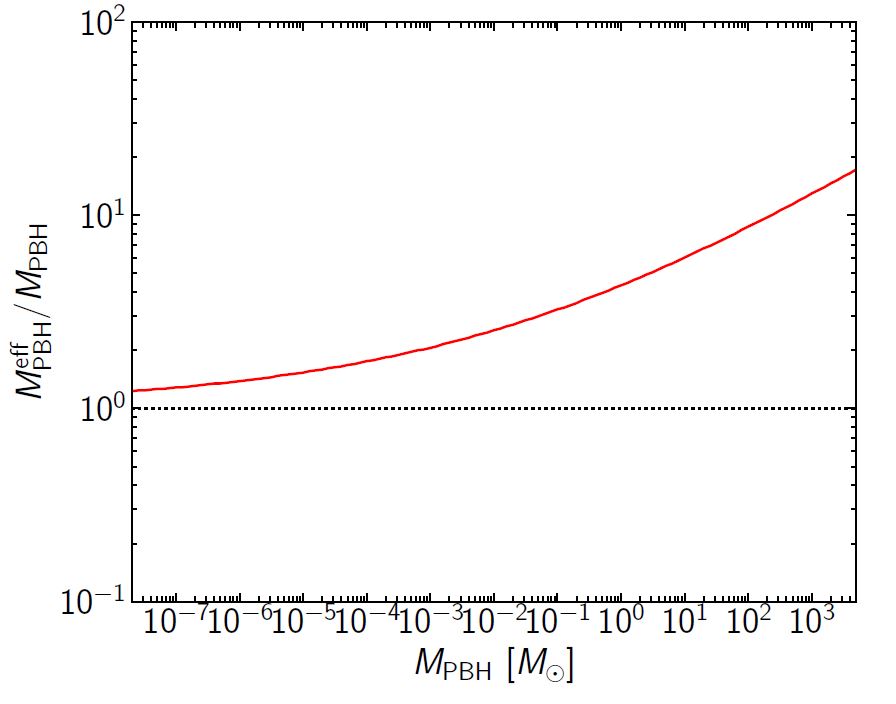}
    \hspace{5mm}
    \caption{\textbf{Left:} The ratio of the effective PBH mass to the PBH mass for microlensing of dressed PBHs for local sources at M31 Andromeda galaxy  \textbf{Right:} Similar to the left panel, but for cosmological sources, assuming $z_s = 1$ and $z_L = 0.5$.}
    \label{fig:halomeff}
\end{figure*}

\section{Cosmological Lensing of FRBs by PBHs}
\label{sec:lensing}

Below we outline the details of cosmological lensing of FRBs by PBHs that we consider, following Ref.~\cite{Munoz:2016tmg}.

Strong lensing of an FRB by a PBH produces two images. For a PBH lens of mass $M_{\rm PBH}$ located at redshift $z_L$, the lensed images are located at distinct positions 
$r_{\pm} = (l \pm \sqrt{l^2 + 4 r_E^2})/2$, where $l$ is the angular impact parameter and $r_E$ is the angular Einstein radius as before. The images are separated in time with the delay
\begin{equation} \label{eq:deltat}
    \Delta t = 4 G M_{\rm PBH} (1 + z_L) \Big[ \dfrac{y}{2}\sqrt{y^2 + 4} + \log \Big( 
    \dfrac{\sqrt{y^2 + 4}+y}{\sqrt{y^2 + 4}- y}\Big)\Big]~,
\end{equation}
where $y = l/r_E$ is the normalized impact parameter.
The corresponding lensing cross-section for a PBH point lens is given by
\begin{align} \label{eq:xsec}
    \sigma (M_{\rm PBH}, z_L) =&~ \dfrac{4 \pi G M_{\rm PBH} D_{\rm ol} D_{\rm ls}}{D_{\rm os}} [y_{\rm max}^2 (\bar{R}_f) - y_{\rm min}^2(M_{\rm PBH}, z_L)]~,
\end{align}
where $y_{\rm min}$ and $y_{\rm max}$ set the range of the considered impact parameters.

The flux ratio denoting absolute value of the ratio of the magnifications $\mu_{\pm}$ of the two lensed images can be defined as
\begin{equation}
    R_f = \Big|
    \dfrac{\mu_+}{\mu_-}\Big| = \dfrac{y^2 + 2 + y \sqrt{y^2 + 4}}{y^2 + 2 - y \sqrt{y^2 + 4}}~.
\end{equation}
Requiring that both images are distinctly observed and none too dim by constraining $R_f$ to be smaller than a critical value $\bar{R}_f$, the maximum impact parameter is constrained to be
\begin{equation}
    y_{\rm max} = \Big(\dfrac{1 + \bar{R}_f}{\sqrt{\bar{R}}_f} - 2\Big)^{1/2}~.
\end{equation}
As in Ref.~\cite{Munoz:2016tmg}, we consider $\bar{R}_f = 5$ throughout our analysis.

To distinguish lensed images as double peaks, separated in time, imposes that $\Delta t$ is larger than the typical millisecond temporal pulse width of observed FRB signal~\cite{CHIMEFRB:2021srp}. Throughout the paper, we require that time delay is larger than critical value $\Delta \bar{t} = 1$~ms. This restriction results in a value of $y_{\rm min}$, which we find by numerically solving Eq.~\eqref{eq:deltat}.

Provided lensing cross-section of Eq.~\eqref{eq:xsec}, for a given FRB at distance $z_S$, the lensing optical depth for a (bare) PBH is
\begin{equation} \label{eq:tau}
\tau^{\rm w/o\,h}(M_{\rm PBH}) = \int_0^{z_S}d\chi (z_L)(1+z_L)^2 n_{\rm PBH}(M_{\rm PBH}) \sigma (M_{\rm PBH}, z_L)~,
\end{equation}
where $\chi(z)$ is the comoving distance at redshift $z$, $n_{\rm PBH} = f_{\rm PBH} \Omega_{\rm cdm} \rho_{\rm crit}/M_{\rm PBH}$ is the comoving number density of PBHs, with $f_{\rm PBH}$ being the fraction PBH DM abundance, $\Omega_{\rm cdm} = 0.24$ is the Universe's cold DM density. 

In order to calculate integrated lensing probability with optical depth, we account for FRB population distribution across redshifts. Following Ref.~\cite{Munoz:2016tmg}, we consider FRBs distributed with a constant comoving number density as
\begin{equation}
N(z) = \mathcal{N} \dfrac{\chi^2(z)}{H(z)(1+z)} e^{-d_L^2(z)/[2 d_L^2(z_{\rm cut})]}~,
\end{equation}
where $d_L$ is luminosity distance, $H(z)$ is the Hubble parameter at redshift $z$, $\mathcal{N}$ is normalization factor for integration of $N(z)$ to unity. We take the redshift model Gaussian cut-off $z_{\rm cut} = 0.5$, which represents potential instrumental signal-to-noise threshold.
 
\bibliography{halopbh.bib}

\end{document}